\documentstyle[aps,psfig,amssymb,bbm]{revtex}

\newcommand{\ket}[1]{\, | #1 \rangle}

\newcommand{\Ge}{\Gamma}

\begin{document}

\draft
\title{Decay rates and survival probabilities in open quantum systems}
\author{Sandro Wimberger,$^{(a,b)}$ Andreas Krug,$^{(a)}$ and Andreas Buchleitner$^{(a)}$} 
\address{$^{(a)}$Max-Planck-Institut f\"ur Physik komplexer Systeme, 
N\"othnitzer Str. 38, D-01187 Dresden \\
$^{(b)}$Universit\`a degli Studi dell' Insubria, Via Valleggio 11, 
I-22100 Como
} 

\date{\today}
\narrowtext
\twocolumn
\maketitle

\begin{abstract}
We provide the first statistical analysis of the decay rates of strongly
driven 3D atomic Rydberg states. The distribution of the rates 
exhibits universal features due to Anderson localization,
while universality of the time dependent decay requires particular 
initial conditions.
\end{abstract}
\pacs{PACS numbers: 05.45.Mt, 32.80Rm, 42.50Hz, 72.15Rn}

%%%%%%%%%%%%% Introduction %%%%%%%%%%%%%%%%%%%%%%%
The macroscopic transport properties of classical and quantum systems
sensitively depend on the dynamics of their microscopic constituents. Whereas
regular Hamiltonian dynamics implies the quasiperiodic confinement of phase
space density, chaotic motion and disorder generically induce diffusion-like
probability transport exploring all accessible phase space at sufficiently
long times. In low-dimensional, disordered quantum systems {\em Anderson
localization} 
efficiently inhibits diffusive transport through quantum
interference, as first predicted by Anderson for the charge transfer across
disordered solid state samples. Later on it was realized that chaos can
substitute for disorder in Anderson's scenario, leading to {\em dynamical
localization} \cite{FGP82,CGS88} 
of quantum probability transport. However, as a further 
complication, quantum systems with classically non-integrable Hamiltonian
dynamics (or, in the absence of a clear classical analog, with
strongly coupled degrees of freedom) most often do not feature purely
chaotic 
but rather mixed regular chaotic
dynamics: besides the chaotic component of phase space 
the latter also comprises regular regions
separated from the chaotic domain by fractal structures like cantori
\cite{mackay84}. Hence,
dynamically localized quantum transport may be amended by quantum tunneling
from regular regions or by semiclassical localization in the vicinity of
partial phase space barriers \cite{geisel86}. 

A rather robust measure of transport is the transmission or decay probability
of some initial probability distribution across or from a confined region
$\Omega $ of
phase space. Whereas the transmission problem immediately suggests a
scattering approach, decay is most conveniently described by the norm
$\parallel {\mathcal P}_{\Omega}\cdot U\ket{\phi_0}\parallel$ of the
time evolved initial state $\ket{\phi_0}$ upon projection 
${\mathcal P}_{\Omega}$ 
on the phase space
domain we want to characterize. 
If $\Omega$ is confined by an absorbing boundary, attached to some leads which
allow for transport to infinity, or coupled to some continuum of states, then
$\parallel {\mathcal P}_{\Omega}\cdot U\ket{\phi_0}\parallel$ will decrease
monotonically in time, for generic $\ket{\phi_0}$
\cite{reed80,abu93,benenti00,steinbach00}. Given the spectral
decomposition of the Hamiltonian, this finite leakage from $\Omega$ can be
accounted for by associating non-vanishing
(exponential) decay rates $\Gamma_j$ with eigenstates $\ket{\epsilon_j}$
with non-vanishing support on $\Omega$. Given some initial
$\ket{\phi_0}={\mathcal P}_{\Omega}\ket{\phi_0}$, 
the survival probability within the domain $\Omega$
boils down to \cite{abu93}
\begin{equation}
P_{\rm surv}(t)=
\parallel {\mathcal P}_{\Omega}\cdot U\ket{\phi_0}\parallel =
\sum_jw_j\exp
(-\Gamma_j t)\; .
\label{psurv}
\end{equation}
All information on the decay process is now encoded \cite{BD95b}
(i) in the set of decay rates -- which is a {\em global} spectral 
property of the problem -- and 
(ii) in the expansion coefficients
$w_j=|\langle\phi_0|\epsilon\rangle|^2$  of the initial wave
packet in the eigenbasis, what is a specific representation of the {\em
initial condition}, and a {\em local} property to the extent that
$\ket{\phi_0}$ is localized.
With these premises, our
initial assertion on the crucial importance of microscopic dynamics for
macroscopic transport leads to the guiding
question of our present contribution \cite{BD95b,ketzmerick00}: 
How do classically mixed
regular-dynamics affect these global and local spectral properties, and,
consequently, the time dependence of $P_{\rm surv}$? Whilst, recently,  
a considerable
corpus of literature has addressed this problem within the context of
simple models of mesoscopic transport \cite{benenti00,ketzmerick00,TG00}, 
we shall here underpin our rather
general answer by a numerically exact treatment of highly excited
three dimensional Rydberg states under strong microwave 
driving \cite{koch95}. These are 
paradigmatic objects to (theoretically and experimentally) 
study \cite{abu93,BD95b,koch95,ak01} quantum
probability decay in the presence of tunneling as well as semiclassical and
dynamical localization, and allow a natural and clean definition of 
$P_{\rm surv}$ and the associated projection ${\mathcal P}_{\Omega}$, without any
approximations. 

Let us start with a brief reminder of dynamical localization in periodically
driven hydrogen Rydberg atoms, to establish the analogy between the general
transport problem depicted above and our specific choice of its physical
realization: The atoms are initially prepared in an unperturbed atomic
eigenstate $\ket{\phi_0}=\ket{n_0\ \ell_0\ m_0}$ labeled by the principal
quantum number $n_0$, the angular momentum $\ell_0$, and the angular momentum
projection $m_0$. For linearly polarized coherent 
driving (which we treat in dipole
approximation) of amplitude $F$ at frequency $\omega$, $m_0$ is a good quantum
number inherited from the rotational symmetry of the problem, and the atomic
excitation process proceeds through the subsequent absorption and/or emission
of photons from/into the field, along a sequence of near resonantly coupled
states $\ket{n\ \ell}$. Provided $F$ is large enough compared to the
attractive force exerted on the electron by the nucleus, the classical
dynamics of the driven Rydberg electron initially launched along a Kepler
orbit turns chaotic, leading to an efficient energy absorption also 
on the quantum
level, which finally leads to the ionization of the atom. 
Hence, we define the electronic 
survival probability as the remaining bound state
population after an interaction time $t$, with ${\mathcal P}_{\Omega}$ 
the projector on the
bound state component of Hilbert space. The $\Gamma_j$ in Eq.~(\ref{psurv})
then become the ionization rates of the Floquet eigenstates $\ket{\epsilon_j}$
of the atom in the field, and the $\omega_j$ the associated 
overlaps (averaged over one field cycle \cite{abu93}) 
of $\ket{\phi_0}$ with $\ket{\epsilon_j}$. 
With some theoretical and
numerical effort, all these quantities can be extracted from the fundamental
Floquet eigenvalue problem, for typical parameter values of state of the art
laboratory experiments \cite{ak01}. 
\begin{figure}
\centerline{\psfig{figure=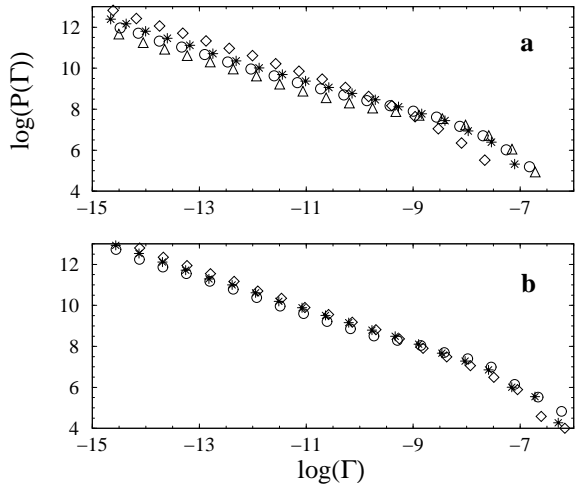,width=7.5cm,angle=270}}
\caption{Distribution of the ionization rates $\Ge$ of
microwave-driven (a) 1D and (b) 3D $m_0=0$ Rydberg states
of atomic hydrogen, for three different values of the localization parameter
${\cal L}= 0.25$ (diamonds), $0.5$ (stars), $1.0$ (circles),
and $2.0$ (pyramids; only in (a), because of the high cost of 3D calculations
at the corresponding high field amplitudes). 
The distributions were generated, at fixed ${\cal L}$, by sampling
the Floquet spectra over the frequency (and corresponding amplitude, $F$) 
ranges $\omega/2\pi =13.16\ldots16.45\
\rm GHz$ (1D) and $35.5\ldots 36.1\ \rm GHz$ (3D), within 
a Floquet zone of width $\omega$ centered around $n_0=70$ (3D) and $100$
(1D), respectively. 
} 
\label{fig1}
\end{figure}

The atomic ionization process is mapped on the Anderson model through the
atomic localization parameter ${\cal L}=\xi/\eta $ \cite{CGS88}, with 
$\xi$ the localization length in units of the photon energy. $\eta$ is the
ionization potential of $\ket{\phi_0}$, measured in units of $\omega$, and is
completely determined by $n_0$, due to the angular momentum degeneracy of the
hydrogen spectrum. Provided that $n_0$ -- through its correspondence with the 
principal action of the classical
dynamics -- is chosen within the chaotic component of phase space 
(possibly garnished by nonlinear
resonance islands and remnants of classically regular motion immerged in the
``chaotic sea''), $\eta$ can be conceived as a measure of the
extension of the domain of complex transport along the
energy axis. If we identify $\eta$ with the sample length in the 
original Anderson
problem and average over different realizations of a
fixed value of ${\cal L}\ll 1$, quantum suppression of 
transport through exponentially localized
eigenfunctions over the quasi-resonantly coupled states on the energy 
axis \cite{CGS88,benenti00,CMS99} implies
a
distribution $P(\Gamma)\sim 1/\Gamma$ of the ionization rates entering
Eq.~(\ref{psurv}), in accord with the current theory \cite{TG00}
on open Anderson insulators.

Statistically
independent realizations of a fixed localization parameter 
can be generated by simultaneous variation of $F$
and $\omega$, due to the relation 
${\cal L}\simeq 6.66 F^2n_0^2/\omega^{7/3}$
from the original theory on dynamical localization
\cite{CGS88} in
periodically driven, one dimensional hydrogen atoms. Whilst this theory has no
quantitative predictive power \cite{abu93,koch95}, 
it is nonetheless known to give qualitatively
correct predictions, even for 3D Rydberg states with not too large values of
the angular momentum quantum number $\ell$. In the sequel we therefore use
this quantity such as to guide our choice of $F$ and $\omega $ values. 

Fig.~\ref{fig1} shows the distribution of ionization rates $\Gamma$ of real 3D
hydrogen Rydberg states and of a one dimensional model atom (where the
Rydberg electron is confined to 1D configuration space $z>0$ defined by
the polarization axis, with the Coulomb singularity at the 
origin \cite{abu93,BD95b}) exposed to
a microwave field, within a Floquet zone (of width $\omega $ on the energy
axis \cite{abu93,ak01}) centered around
the $n_0=70$ (3D) and $n_0=100$ (1D) Rydberg manifold. 
The different values of ${\cal L}=0.25\ldots 2.0$ are realized
by sampling the spectra over frequency ranges $\omega/2\pi =35.5\ldots 36.1\
\rm GHz$ (3D), and $13.16\ldots 16.45\ \rm GHz$ (1D),
respectively, and adjusting $F$ correspondingly. 
Due to the dramatically enhanced spectral density of the 3D as compared
to the 1D problem (a consequence of the additional degree of freedom labeled
by $\ell$) only ten equidistant $\omega$-values are needed to generate the
3D data (with a total sample size of approx. 25000 eigenstates),
in contrast
to 500 in the 1D case (up to 100000 eigenvalues). 

For small localization
parameter ${\cal L}=0.25$, both, the 1D 
model atom as well as the real object do
exhibit ionization rates distributed according to the $\Gamma ^{-1}$ law, over
approx. seven orders of magnitude from $\Gamma\simeq 10^{-15}$ to
$\Gamma\simeq 10^{-9} \ \rm a.u.$, in nice 
agreement with the prediction for decay from a
disordered solid \cite{TG00}. 
However, as the localization parameter is increased (by
systematically increasing $F$ over the entire frequency range indicated above)
to ${\cal L}=1.0$, 
we observe a depletion of the distribution at small
rates, balanced by a decrease of the slope of $P(\Gamma)$ in an intermediate
range $\Gamma\simeq 10^{-10}\ldots 10^{-8}\ \rm a.u.$ Such behavior is
clearly incompatible with the simple picture of exponentially localized
probability densities along the energy axis (due to dynamical localization of
the eigenstates over the chaotic phase space component)
\cite{CGS88,benenti00,CMS99},  
and suggests the
presence of alternative transport mechanisms which compete with dynamically
localized diffusion \cite{BD95b,ketzmerick00}. 
As already anticipated in the introduction, such
observation does not come as a surprise, since periodically driven Rydberg
states exhibit a mixed regular chaotic phase space. Indeed, it is known from
experimental
as well as exact theoretical/numerical studies
\cite{BD95b,koch95,delande91} of strongly
perturbed atomic Rydberg systems, and, more recently, from mesoscopic systems
with decay \cite{ketzmerick00}, that the eigenstates 
of such quantum systems can be classified 
according to their localization \cite{BD95b,ketzmerick00,richards94} 
(i) on regular and/or elliptic regions, (ii)
along remnants of regular motion immerged in the chaotic sea
(``separatrix/hierarchical states'' living on ``can-tori''), 
and (iii) in the chaotic domain of
phase space. Furthermore, elliptic regions as well as 
remnants of invariant
tori are rather robust under changes of some control parameter such as $F$ in
our present case \cite{BD95b}. 
\begin{figure}
\centerline{\psfig{figure=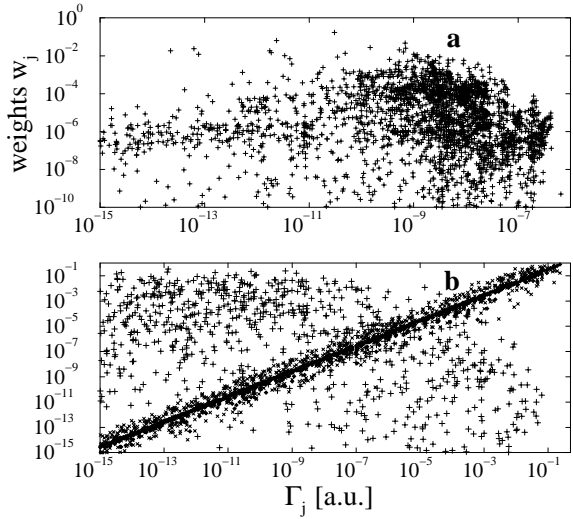,width=7.5cm,angle=270}}
\caption{(a) Expansion coefficients $w_j$ of the 3D atomic initial state 
$\ket{n_0=70, \ell_0 =0, m_0=0}$ in the Floquet eigenstates
$\ket{\epsilon_j}$, for ${\cal L}=1.0$ and $\omega/2\pi =
35.6\ \rm GHz$, vs. the associated decay rates $\Gamma_j$. 
(Qualitatively similar results are obtained for the
one dimensional model atom, as well as for microwave driven alkali Rydberg
states \protect\cite{Kru01T}.)
(b) Same as (a), but for initial states prepared at the sites  
$900$ (plusses), $999$ (crosses), $1000$ 
(diamonds clustered along the diagonal) of a 1D
Anderson model of sample length $N=1000$ 
(uniformly distributed on-site potentials $V_{n_i}$ with
$|V_{n_i}|\leq 1.5$, constant coupling $V_{n_i,n_i\pm1}=1$, and an absorbing
boundary at one end, introduced by adding an imaginary part $-i\gamma
=-i\times 0.31$ to the on-site potential at $n_i=1000$; results for 10
realizations of the potential are gathered together).
%%%% SAW
%%%%of sample length $N=1000$. 
Only in the special case of the 
initial state placed right at the edge $n_{\rm i}=1000$ 
of the sample is there a one-to-one
relation between decay rates and expansion coefficients!
} 
\label{fig2}
\end{figure}
Accordingly, the localization properties as well as the
decay rates of the associated eigenstates tend to be less sensitive when
changing $F$, whereas those states localized in the chaotic domain, with their
decay rates essentially determined by $\cal L$, exhibit rapidly increasing
rates as $F$ is increased. Hence, we attribute the deviation from the $\Gamma
^{-1}$ law at large values of $F$ in Fig.~\ref{fig1} to the rapid increase of
the decay rates of class (iii) states with $\cal L$ (corresponding to the
effective destruction of dynamical localization), whereas class (i) and (ii)
states accumulate at intermediate $\Gamma$-values at the largest $\cal
L$-values considered here (we verified this assertion by inspection of the 
Husimi phase space representations of the 
Floquet eigenstates of the 1D problem \cite{BD95b}). 
At small values of $\cal L$, the latter exhibit
extremely small ionization rates below $\Gamma\simeq 10^{-13}\ \rm a.u.$ (or
even below $\Gamma\simeq 10^{-15}\ \rm a.u.$, 
this is below our numerical precision, hence the plot
does not extend into that domain), what leads to isolated, extremely
narrow resonances in experimental or numerical photoionization cross 
sections \cite{BD95b,delande91}.
Therefore, in this parameter regime, $P(\Gamma)$ is dominated by the 
decay rates of the dynamically localized class (iii) states, in agreement with
the Anderson scenario. 
Let us also comment on the apparently distinct weight of class (i) and
(ii) states in the 1D and the 3D situation, which is born out in
Fig.~\ref{fig1}: the depletion of the low-$\Gamma$ part is less pronounced for
the real atom, and equally so the region of decreased slope at intermediate
$\Gamma$-values. This is due to the contribution of high $\ell$ states (at
fixed $m_0$) to the rate distribution -- unavailable in 1D -- which exhibit 
vanishing decay rates and do not contribute to the plotted distribution at
small $\cal L$. As
$\cal L$ is increased, these states contribute to $P(\Gamma)$ in the range
$\Gamma < 10^{-11}\ \rm a.u.$, such as to
compensate for the re-shuffling of low-$\ell$ states from small to intermediate
$\Gamma$'s. This smooths the 3D distribution as compared to the 1D result.

Hence, we have seen how chaotic transport can mimic signatures of disorder in
the quantum decay rate 
distribution, and induce amendments thereof due to peculiar
structures of mixed regular chaotic dynamics. As obvious from
Eq.~(\ref{psurv}), $P(\Gamma)$ 
%%%%% SAW
%%nonetheless 
does not determine the
(experimentally often very easily accessible \cite{abu93,BD95b,ak01}) 
survival probability alone, but
only in conjunction with the expansion coefficients $w_j$. Starting from a
general form $P(\Gamma)\sim\Gamma^{-\alpha}$, many authors infer a
``universal'' time dependence $P_{\rm surv}\sim 1/t^{2-\alpha}$, leaning on  
the
additional assumption $w_j\sim\Gamma_j$ \cite{benenti00,steinbach00}. 
The latter is nonetheless {\em only}
justified if the initial state $\ket{\phi_0}$, which represents a {\em local}
probe of the spectrum, is localized close to the boundary of $\Omega$ -- which
is the ``edge of the sample'' in
a language adapted to the Anderson scenario, and the 
ionization threshold
$n\rightarrow\infty$ in the atomic problem. 
For a generic choice of $\ket{\phi_0}$ ``within the sample'', i.e. somewhere
within $\Omega$, this basic assumption is utterly misleading, as
illustrated in Fig.~\ref{fig2}. In Fig.~\ref{fig2} a), we show the
distribution of the $w_j$ and of the corresponding $\Gamma_j$, on a {\em double
logarithmic} plot, for $\ket{\phi_0}=\ket{n_0=70\ \ell_0=0}$ in driven
3D hydrogen. The essentially uncorrelated $w_j$ and $\Gamma_j$ nicely
correlate with the earlier finding that phase space localization properties
and decay rates of strongly driven Rydberg atoms are {\em not} unambiguously 
related \cite{BD95b}, and clearly devalidate the above proportionality 
assumption. 
On
the other hand, it can easily be shown within a simple, one dimensional
Anderson model how the generic situation of Fig.~\ref{fig2} a) continuously
evolves into the situation of strongly correlated $w_j$ and $\Gamma_j$, as
illustrated in Fig.~\ref{fig2} b): There, we calculated the decay rates of the
eigenstates of a 1D tight binding Hamiltonian of length $N=1000$, with
diagonal disorder and constant off-diagonal terms. Decay is introduced in this
model by absorbing boundary conditions at one end, whereas a reflecting wall
is imposed at the other. If $\ket{\phi_0}$ is placed well inside the sample,
expansion coefficients and decay rates are uncorrelated, though the
distribution collapses onto a straight line $w_j\sim\Gamma_j$ as
$\ket{\phi_0}$ is shifted towards the open end (site $n_{\rm i}=1000$)  
of the sample. It is this
latter situation which was realized in recent numerical calculations on
periodically driven hydrogen atoms additionally subjected to an intense static
electric field \cite{benenti00}. 
\begin{figure}
\centerline{\psfig{figure=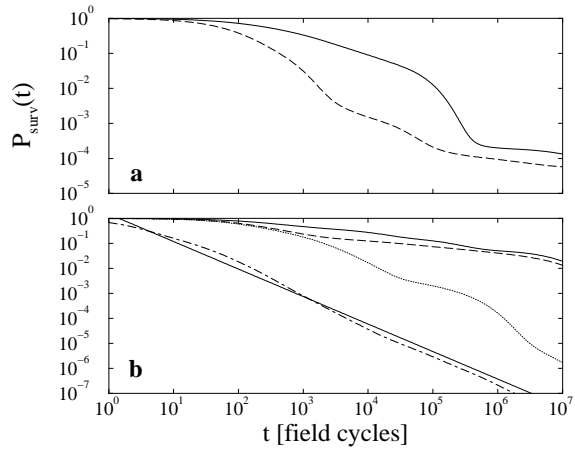,width=7.5cm,angle=270}}
\caption{Survival probability $P_{\rm surv}(t)$
obtained from distributions as in Figs.~\ref{fig1} and 
\ref{fig2}
via Eq.~(\ref{psurv}). (a) 1D hydrogen, initial state $\ket{n_0=100}$, 
driving
frequencies and localization parameters $\omega/2\pi = 16.45\ \rm GHz$, ${\cal
L}=1.0$ (full line), $13.16\ \rm GHz$ and $2.0$ (dashed); 
(b) 3D hydrogen, $n_0=70$, $m_0=0$, $\ell_0=0$
(solid line), $15$ (dashed), $45$ (dotted), $\omega/2\pi = 35.6\ \rm GHz$, 
${\cal L}=1.0$. In this latter case, different
$w_j$-distributions are realized by changing the angular momentum
$\ell_0$ of the initial atomic state, always leading to 
near-algebraic decay (like in 1D), though with
distinct (and {\em non-universal}) decay exponents.
Only fictitious 
expansion coefficients 
$w_j \equiv \Gamma_j/<\Gamma>$ induce asymptotically ``universal'' 
decay 
$P_{\rm surv}(t)\sim t^{- \gamma}$ (dash-dotted line),
$\gamma\simeq 2- \alpha \simeq 1.1$ (indicated by the full line), 
with $\Gamma_j$, $<\Gamma>$, and
$\alpha\simeq 0.9$ extracted from the $\Gamma<10^{-10}\ \rm a.u.$ part of the 
${\cal L}=1.0$ data
in Fig.~\ref{fig1}(b). 
} 
\label{fig3}
\end{figure}

Given the results illustrated in Fig.~\ref{fig2}, we can no more hope for a
universal decay law for the survival probability $P_{\rm surv}(t)$ -- despite
the rather universal features of $P(\Gamma)$. This is finally illustrated in
Fig.~\ref{fig3}, where we show $P_{\rm surv}(t)$ for 1D and 3D hydrogen in the
case of large $\cal L$: Whilst algebraic decay of $P_{\rm
surv}(t)\sim t^{-\gamma}$ is generic (even if modulated by local
fluctuations \cite{saw01}), 
simply due to the broad distribution of the $w_j$ and
$\Gamma_j$ as illustrated in Figs.~\ref{fig1} and \ref{fig2}, the
``universal'' decay exponent $\gamma =2-\alpha$ is only observed in accidental
cases -- or if we reshuffle the expansion coefficients of Fig.~\ref{fig2} a)
according to $w_j\rightarrow\Gamma_j/\langle \Gamma \rangle $, with 
$\langle\Gamma\rangle $ the average rate over the entire spectrum, as
illustrated in Fig.~\ref{fig3} b). In this, however, unrealistic case, 
$P_{\rm surv}(t)$ asymptotically decays with 
$\gamma\simeq 1.1$, which perfectly matches the
prediction $\gamma =2-\alpha$, with $\alpha =0.9$ extracted from the
small $\Gamma$ ($<10^{-10}\ \rm a.u.$) part of the 
corresponding data set in Fig.~\ref{fig1} b). Therefore, universal features of
the statistics of the decay rates of classically chaotic quantum systems do
generically {\em not} carry over to the time dependent decay of the survival
probability of some initial state prepared at an arbitrary location in phase
space. This simply expresses the essential decorrelation of the phase space
localization properties of the eigenstates of chaotic quantum systems and of
their asymptotics -- which {\em alone} determine their decay. 

%%%%%%%%%%%%%%% \acknowledgments %%%%%%%%%%%%%%%%%%%%%%%%%%%%%%
Support from EU Program QTRANS RTN1-1999-08400, 
is gratefully acknowledged (S.W.).

%%%%%%%%%%%% change for submission!!! %%%%%%%%%%%%%%%%%%
%\begin{references}
\bibliographystyle{abuprsty}
\bibliography{bibabu,bibliobook}
%\end{references}

%%%%%%%%%%%%%%%%%%%%%% Figues: #=1 %%%%%%%%%%%%%%%%%%%%%%%%%%%%%%%%%%%%% 

\end{document}